\documentclass[11pt,preprint]{aastex}

\usepackage{epstopdf}

\newcommand{\crra}{Cram\'er-Rao}

\slugcomment{Accepted by PASP}

\shorttitle{\crra\ in joint astrometric and photometric estimation}
\shortauthors{Mendez et al.}

\begin{document}

\title{Analysis of the \crra\ lower uncertainty bound in the joint
  estimation of astrometry and photometry}

\author{Rene A. Mendez\altaffilmark{1}}

\affil{Departamento de Astronom\'{\i}a, Facultad de Ciencias
  F\'{\i}sicas y Matem\'aticas,\\Universidad de Chile, Casilla 36-D,
  Santiago, Chile} \email{rmendez@u.uchile.cl}

\author{Jorge F. Silva}

\affil{Department of Electrical Engineering, Av. Tupper 2007,
  Santiago, Chile \\Information and Decision System
  Group\\ Universidad de Chile} \email{josilva@ing.uchile.cl}

\author{Rodrigo Orostica}

\affil{Departamento de Astronom\'{\i}a, Facultad de Ciencias
  F\'{\i}sicas y Matem\'aticas,\\Universidad de Chile, Casilla 36-D,
  Santiago, Chile} \email{rorostica@ing.uchile.cl}

\author{and}

\author{Rodrigo Lobos}

\affil{Department of Electrical Engineering, Av. Tupper 2007,
  Santiago, Chile \\Information and Decision System
  Group\\ Universidad de Chile} \email{rlobos@ing.uchile.cl}

\altaffiltext{1}{On leave at the European Southern Observatory,
  Casilla 19001, Santiago, Chile.}

\begin{abstract}
In this paper we use the \crra\ 
lower uncertainty bound to estimate the maximum precision that could
be achieved on the joint simultaneous (or 2D) estimation of photometry
and astrometry of a point source measured by a linear CCD detector
array. We develop exact expressions for the Fisher matrix elements
required to compute the \crra\ bound in the case of a source with a
Gaussian light profile. From these expressions we predict the behavior
of the \crra\ astrometric and photometric precision as a function of
the signal and the noise of the observations, and compare them to
actual observations - finding a good correspondence between them.

From the \crra\ bound we obtain
the well known fact that the uncertainty in flux on a Poisson-driven
detector, such as a CCD, goes approximately as the square root of the
flux. However, more generally, higher order correction factors that
depend on the ratio $B/F$ or $F/B$ (where $B$ is the background flux
per pixel and $F$ is the total flux of the source), as well as on the
properties of the detector (pixel size) and the source (width of the
light profile), are required for a proper calculation of the minimum
expected uncertainty bound in flux. Overall the \crra\ bound predicts
that the uncertainty in magnitude goes as $(S/N)^{-1}$ under a broad
range of circumstances.

As for the astrometry we show that its \crra\ bound also goes as
$(S/N)^{-1}$ but, additionally, we find that this bound is quite
sensitive to the value of the background - suppressing the background
can greatly enhance the astrometric accuracy.

We present a systematic analysis of the elements of the Fisher matrix
in the case when the detector adequately samples the source
(oversampling regime), leading to closed-form analytical expressions
for the \crra\ bound. We show that, in this regime, the joint
parametric determination of photometry and astrometry for the source
become decoupled from each other, and furthermore, it is possible to
write down expressions (approximate to first order in the small
quantities $F/B$ or $B/F$) for the expected minimum uncertainty in
flux and position. These expressions
are shown to be quite resilient to the oversampling condition, and
become thus very valuable benchmark tools to estimate the approximate
behavior of the maximum photometric and astrometric precision
attainable under pre-specified observing conditions and detector
properties.
\end{abstract}



\newtheorem{proposition}{\bf Proposition}

\section{Introduction}
\label{sec_intro}
In this paper we extend the 1D \crra\ analysis done in \citet{men13}
to the 2D case of simultaneous photometry and astrometry estimation on
a linear CCD detector. The goal is to provide an estimation setting
that is more realistic than that presented in \citet{men13}, while
still being tractable analytically so that useful closed-form
expressions can be derived and interpreted from the analysis. This
scenario allows us also to explore, in a simple manner, the extent of
the inter-dependence between astrometry and photometry, from the point
of view of the
\crra\ error bound under different instrumental and detection regimes.

In general, the \crra\ lower variance bound can be used to cover a
broad span of applications, ranging from instrument design for
specific target accuracy goals, to observational planning, and to data
analysis benchmarking (see., e.g., \citet{peet89}, \citet{jaet92},
\citet{zaet95}, \citet{ad96}). For example, the \crra\ bound can be
used to predict how a particular design choice (pixel size, readout
noise, etc.) influences the photometric and astrometric performance of
the planned instrument, it permits the prediction of lower bounds to
photometric errors for point sources (and for surface photometry of
extended objects), and places lower bounds to the precision with which
the position of point sources can be measured (depending on their
shape), be it as isolated objects, or in a cluster. The
\crra\ formalism also allows us to determine the influence of
sub-pixel dither patterns on the astrometric and photometric errors
(\citet{men13} and this paper,
Section~\ref{subsec_Fisher_int}). Finally, the \crra\ lower bound can
be used to test the statistical adequacy of different data reduction
and analysis algorithms, or even the reliability of our data: Those
pipelines that can not attain the \crra\ bound may not be
statistically optimum.

One of the limitations of the \crra\ formalism is that, in general, by
itself, it does not offer a way to construct an estimator that reaches
the bound
(unless the parametric setting satisfies a necessary and sufficient
condition, see \citet[p.12 ]{Ord04}). However, what one can do is to
try various estimators, in a more or less heuristic way, and compare
its empirical performance, in terms of its variance, with that
predicted by the \crra\ bound, to determine how close it approaches
the bound. An important point to note here is that a biased estimator
may have a variance lower than that predicted by the \crra\ bound (for
a nice and simple example of this see \citet{stpe90}). Therefore, a
very tight estimator should be viewed with caution, since it may be
indicative that our {\it estimations} are actually biased, rendering
parameter estimations that suffer from a systematic effect.

Our paper is organized as follows: Section \ref{sec_pre} introduces
the basic setting of problem, its notations and basic terminology and
results. In addition this section focuses on the simple 1D case of
photometric estimation, and revisits key results of the 1D astrometric
problem.  Section~\ref{main_section} is the main section and
elaborates and analyzes the expression of the \crra\ bound for the
joint astrometry and photometry estimation problem. Finally in
Section~\ref{conc} we summarize our main conclusions.

\section{Preliminaries}
\label{sec_pre}
In this Section we introduce our notation and provide the basic setting
that will be used in the joint astrometric and photometric estimation
problem in Section \ref{main_section}.

\subsection{Parameter estimation and the multivariate \crra\ bound}

Let $I_i$ (with $i=1,...,n$) be a collection of independent
observations (or measurements) that follow a parametric probability
mass function $f_{\vec{\theta}}$ defined on $\mathbb{N}$. The
parameters to be estimated from the measurements $\vec{I}=\left\{
I_i:i=1,...,n\right\}$ will be denoted by
$\vec{\theta}=(\theta_1,\theta_2,...,\theta_m) \in \mathbb{R}^m$. Then
given the measurements, let us consider
$\hat{\vec{\theta}}(I_1,...,I_n)=
(\hat{\theta}_1,\hat{\theta}_2,...,\hat{\theta}_m)$ to be an unbiased
estimator of the parameters $\vec{\theta}$. If
$L(\vec{I};\vec{\theta})$ is the likelihood of the observations
$\vec{I}$ given the parameters $\vec{\theta}$, and we can verify that
$L(\vec{I};\vec{\theta})$ satisfies the condition:
\begin{equation} \label{cond2d}
\mathbb{E}_{\vec{I} \sim f^n_{\vec{\theta}}}\left( \frac{\partial \ln
  L(\vec{I}; \vec{\theta}) }{\partial \theta_i} \right) = 0 \;\;\;
\forall \, \theta_i \;\; (i=1...m)
\end{equation}
then, the celebrated  \crra\ bound states that \citep{ra45,cr46}:
\begin{equation}\label{varcr}
Var (\hat \theta_i(\vec{I})) \geq \sigma^2_{\theta_i} \equiv [
  \mathcal{I}_{\vec{\theta}}(n)^{-1} ]_{i,i}
\end{equation}
where $\mathcal{I}_{\vec{\theta}}(n)$ denotes the {\em Fisher
  information} matrix of the data about the vector of parameters
$\vec{\theta}$, given by:
\begin{equation}\label{fisher}
[ \mathcal{I}_{\vec{\theta}}(n)]_{i,j} = \mathbb{E} \left(
\frac{\partial \ln L(\vec{I}; \vec{\theta})}{\partial \theta_i} \cdot
\frac{\partial \ln L(\vec{I}; \vec{\theta}) }{\partial \theta_j}
\right) \;\; (i,j=1...m).
\end{equation}

\subsection{Joint photometric and astrometric estimation setting}
\label{sub_sec_joint_estimation}

Given a point source parameterized by its position $x_c$ and flux
$\tilde{F}$, the central estimation problem here is to jointly
estimate the pair $(x_c, \tilde{F})$ using the measurements of a
photon integrating device with $n$ pixels (such as a CCD). This device
measures the vector $\left\{I_i:i=1,...,n\right\}$ corresponding to
fluxes (counts) per pixel. In this digital setting, we model
$\left\{I_i:i=1,...,n\right\}$ as independent and not identically
distributed random variables, where $I_i$ follow a Poisson
distribution with expected value given by the function $\lambda_i(x_c,
\tilde{F})$. More precisely, the likelihood function of this
estimation problem is given by:
\begin{equation}\label{eq_sub_sec_joint_estimation_1}
L(\vec{I};(x_c,\tilde{F}))=f_{\lambda_1(x_c,\tilde{F})}(I_1) \cdot
f_{\lambda_2(x_c,\tilde{F})}(I_2) \cdots
f_{\lambda_n(x_c,\tilde{F})}(I_n)
\end{equation}
where $f_{\lambda}(I)=\frac{e^{-\lambda}\cdot \lambda^I}{I!}$ is the
Poisson probability mass function. Note that
equation~(\ref{eq_sub_sec_joint_estimation_1}) models the fact that
the measurements are independent but in general not identically
distributed.

If $\tilde{F}_i(x_c, \tilde{F}) $ represents the expected flux from
the source (at pixel $i$, in photo-e$^-$) and $\tilde{B}_i$ is the
total integrated background (at pixel $i$, in e$^-$), the expected
flux at pixel $i$ follows an additive noise model given by:
\begin{equation}\label{eq_sub_sec_joint_estimation_1b}
\lambda_i(x_c, \tilde{F})=\tilde{F}_i(x_c, \tilde{F}) + \tilde{B}_i,
\end{equation}

Note that in equation~(\ref{eq_sub_sec_joint_estimation_1b}),
$\tilde{B}_i$ includes the contribution from the detector (read-out
noise and dark current) and the the sky background\footnote{See the
  concrete expression in \citet[equation~(23)]{men13}.}, and
consequently it is independent of $(x_c, \tilde{F})$.  On the other
hand, $\tilde{F}_i(x_c, \tilde{F})=\tilde{F} \cdot g_i(x_c)$ where
$g_i(x_c)$ is characterized by the one dimensional normalized ``Point
Spread Function'' (PSF hereafter), denoted by $\phi(x)$~arcsec$^{-1}$,
through:
\begin{equation}\label{eq_sub_sec_joint_estimation_2}
g_i(x_c) = \int_{x_i-\frac{\Delta
    x}{2}}^{x_i+\frac{\Delta x}{2}} \Phi(x-x_c) \, dx.  
\end{equation}
In equation~(\ref{eq_sub_sec_joint_estimation_2}), $x_i$ denotes the
central coordinate of pixel $i\in \left\{1,...,n \right\}$, $\Delta x$
is the pixel size and $\int_{-\infty}^{+\infty} \Phi(x) \, dx = 1$.
In this work we will assume a Gaussian PSF, i.e.,
\begin{equation}\label{eq_sub_sec_joint_estimation_3}
\Phi(x) = \frac{1}{\sqrt{2 \pi} \, \sigma} \, e^{-\frac{1}{2} \left(
  \frac{x}{\sigma}\right)^2} \; \; [\mbox{arcsec}^{-1}]\label{psf}
\end{equation}
which is a reasonable assumption in the context of ground-based data
\citep{men10}. Then, from (\ref{eq_sub_sec_joint_estimation_2}), we
have the following identity that will be used in the computation of
the \crra\ bound in (\ref{fisher}): $\forall i\in
\left\{1,...,n\right\}$
\begin{equation}\label{eq_sub_sec_joint_estimation_4}
\frac{d g_i}{d x_c}(x_c) = \frac{1}{\sqrt{2 \pi} \, \sigma} \left(
e^{-\gamma(x^-_i-x_c)} - e^{-\gamma(x^+_i-x_c)} \right) \; \;
[\mbox{arcsec}^{-1}]
\end{equation}
where $\gamma(x) \equiv \frac{1}{2}(\frac{x}{\sigma})^2$, with
$x^{\_}_i = x_i - \frac{\Delta x}{2}$ and $x^+_i= x_i + \frac{\Delta
  x}{2}$.

Finally, we identify $\tilde{F}$ as the total flux of the source:
\begin{equation}\label{eq_sub_sec_joint_estimation_5}
\sum_{i=1}^n \tilde{F}_i(x_c, \tilde{F}) = \tilde{F} \sum_{i=1}^n
g_i(x_c) = \tilde{F} \sum_{i=1}^n \int_{x_i-\frac{\Delta
    x}{2}}^{x_i+\frac{\Delta x}{2}} \Phi(x) \, dx \approx \tilde{F}
\int_{-\infty}^{+\infty} \Phi(x) \ dx = \tilde{F}.
\end{equation} 
where we have assumed that the detector properly samples the PSF.

\subsection{Photometric estimation} \label{phot1d}

In this section we elaborate on the simplified case of estimating the
flux of a source, $\tilde{F}$ (in units of photo-e$^-$), assuming that
$x_c$ is known with very high accuracy. Hence the (expected) source
flux on pixel $i$ can be written in the form:
\begin{equation} \label{fluxn}
\tilde{F}_i (\tilde{F})= \tilde{F} \cdot g_i(x_c) \;\;\; (i=1...n)
\end{equation}
where the positional parameter $x_c$ is a known quantity in this
context. Then we can verify equation~(\ref{cond2d}) and derive the
\crra\ lower bound for the estimation of $\tilde{F}$ as follows:

\begin{proposition}\label{pro_photo_1D}
Let $\hat{\tilde{F}}(\vec{I})$ be an arbitrary unbiased estimator of
$\tilde{F}$, then:
\begin{eqnarray}
Var(\hat{\tilde{F}}(\vec{I}) ) \ge \sigma^2_{{\tilde{F}}_{1D}} & \equiv &
\frac{1}{\displaystyle \sum_{i=1}^{n} \frac{ g_i^2 } { \tilde{F} \cdot
    g_i + \tilde{B}_{i} }} \label{eq_phot1d_1} \\ & = & 2 \pi \sigma^2
\cdot \tilde{B} \cdot \frac{1} {\displaystyle \sum_{i=1}^{n}
  \frac{J_i(x_c)^2}{\left( 1 + \frac{1}{\sqrt{2 \pi} \, \sigma}\cdot
    \frac{\tilde{F}}{\tilde{B}} \cdot J_i(x_c)
    \right)}}, \label{eq_phot1d_2}
	\end{eqnarray}
where for the last expression, the background $\tilde{B}$ is
considered to be uniform across the array\footnote{The analysis to
  correctly characterize the background for computing the \crra\ bound
  in astronomical applications is elaborated in
  \citet[Sec. 4]{men13}.}, i.e., $\tilde{B}_i=\tilde{B}$ for all $i$,
and where $J_i(x_c)$ is given by:
\begin{equation}\label{ji}
J_i(x_c) \equiv \displaystyle \int_{x^{\_}_i}^{x^+_i}
e^{-\gamma(x-x_c)} \, dx \; \; [\mbox{arcsec}]
\end{equation}
(The proof is presented in Appendix \ref{proof_pro_photo_1D}).

\end{proposition}

From equation~(\ref{eq_phot1d_1}) it is straightforward to compute the
two extreme regimes, i.e., background or source-dominated, which are
given, to first order in the small quantity $\tilde{F}/\tilde{B}$ or
$\tilde{B}/\tilde{F}$ respectively, by:
\begin{equation} \label{fluxcrapp}
\sigma^2_{{\tilde{F}}_{1D}} \simeq \left\{
	\begin{array}{cc}
	\frac{\tilde{B}}{ \sum_{i=1}^{n} g_i^2 } \cdot \left( 1 +
        \frac{\tilde{F}}{\tilde{B}} \cdot \frac{ \sum_{i=1}^{n}
          g_i^3}{ \sum_{i=1}^{n} g_i^2} \right) & \mbox{if $\tilde{F}
          \ll \tilde{B}$} \\ \tilde{F} \cdot \left( 1 + n \cdot \frac{
          \tilde{B} }{\tilde{F}} \right) & \mbox{if $\tilde{F} \gg
          \tilde{B}$}
	\end{array}
	\right.
\end{equation}
where have assumed a constant background as a function of position in
the array, $\tilde{B}$, and used the fact that $\sum_{i=1}^{n} g_i =
1$\footnote{Note that, since $\tilde{B}$ is the background {\it per
    pixel}, the term $n \cdot \tilde{B}$ represents the total
  contribution of the background to the measured flux. In this
  context, $n$ represents not the full pixel array but, rather, the
  portion of the array over which the flux of the source is being
  calculated (see definition of $N_{\mbox{pix}}$ on
  equation~(\ref{up}) below). If this is the case, then
  $\sum_{i=1}^{n} g_i$ is not necessarily equal to one, but rather it
  corresponds to the fraction of the flux enclosed within the $n$
  pixels. Hopefully the ``aperture'' is chosen to include most of the
  flux $\tilde{F}$, or a suitable correction is applied (e.g., through
  a curve-of-growth) to compensate for the missing fraction of this
  flux.}. Interestingly, the second relation above shows the well
known fact that the uncertainty in flux goes approximately as the
square root of the flux itself (measured in e$^-$), when the
background is negligible.

\subsection{Astrometric estimation}
\label{sub_sec_astrot1d}

Here we summarize the main results derived in \citet{men13} for the
problem of estimating the position of the source $x_c$ when the total
flux $\tilde{F}$ is known by the observer.  In terms of notation, we
can consider the expected flux at pixel $i$ by
$\lambda_i(x_c)=\tilde{F}_i(x_c) + \tilde{B}_i$ where
$\tilde{F}_i(x_c)=\tilde{F}\cdot g_i(x_c)$ and $\tilde{F}$ is known.
Then we can state the following:

\begin{proposition} (\citet[equations~(10) and~(21)]{men13}) 
	\label{pro_astro_1D}
	Let $\hat{x_c}(\vec{I})$ be an arbitrary unbiased estimator of $x_c$, 
	then:
	\begin{eqnarray}
		Var(\hat{x}_c(\vec{I})) \ge \sigma^2_{x_{c_{1D}}} & \equiv &
                \frac{1}{\displaystyle \sum_{i=1}^{n} \frac{ \left(
                    \tilde{F} \, \frac{d g_i}{d x_c}(x_c) \right)^2}{
                    \tilde{F} \, g_i(x_c) + \tilde{B}_{i}
                }} \label{eq_sub_sec_astrot1d_1} \\
  		& = & 2 \pi \sigma^2 \cdot
                \frac{\tilde{B}}{\tilde{F}^2} \cdot
                \frac{1}{\displaystyle \sum_{i=1}^{n} \frac{\left(
                      e^{-\gamma(x^-_i-x_c)} - e^{-\gamma(x^+_i-x_c)}
                      \right) ^2}{\left( 1 + \frac{1}{\sqrt{2 \pi} \,
                        \sigma}\frac{\tilde{F}}{\tilde{B}} \cdot J_i(x_c) \right)}
                }. \label{eq_sub_sec_astrot1d_2}
	\end{eqnarray}
In the last expression we have assumed a uniform background
$\tilde{B}$ across pixels, as in equation~(\ref{eq_phot1d_2}).

\end{proposition}

In the high resolution regime, i.e., $\Delta x/\sigma \ll 1$, 
the following limiting (weak and strong source) closed-form expression for 
$\sigma^2_{x_{c_{1D}}}$ can be derived (see details in \citet[Sec. 4.1.]{men13}): 

\begin{equation} \label{eq_sub_sec_astrot1d_3}
\sigma^2_{{x}_{c_{1D}}} \approx \left\{
\begin{array}{cc}
\frac{\sqrt{\pi}}{2 \, (2 \, \ln 2)^{3/2}} \cdot \frac{\tilde{B}}{\tilde{F}^2} \cdot
\frac{FWHM^3}{\Delta x} & \mbox{if $\tilde{F} \ll \tilde{B}$} \\
\frac{1}{8 \, \ln 2} \cdot \frac{1}{\tilde{F}} \cdot FWHM^2 & \mbox{if $\tilde{F}
  \gg \tilde{B}$},
\end{array}
\right.
\end{equation}
where $FWHM = 2 \sqrt{2 \ln 2} \,\, \sigma$ denotes the ``Full-Width
at Half-Maximum'' parameter, which is associated with the image
quality at the observing site.

\section{Joint astrometric and photometric \crra\ bound}
\label{main_section}

Let us now consider the more realistic case of having to jointly
estimate the flux $\tilde{F}$ and astrometric position $x_c$ on a
linear detector. Note that the calculation of the inverse Fisher
matrix in equation~(\ref{varcr}) implies computing its determinant,
which, in general, involve all the elements of the matrix. This
property highlights the potential cross-dependency in the errors of
quantities that one may naively consider, in principle, as decoupled,
like, e.g., 1D astrometry and photometry presented in
Section~\ref{sec_pre}.  This will be further explored in
Section~\ref{range}.

From equation~(\ref{eq_sub_sec_joint_estimation_1}) we have that $\ln
L(\vec{I}; (x_c,\tilde{F})) = {\sum_{i=1}^n \left( I_i\cdot \ln
  \lambda_i(x_c,\tilde{F}) - \lambda_i(x_c,\tilde{F}) - \ln I_i!
  \right)}$. In this case, it is straightforward to verify that the
conditions in equation~(\ref{cond2d}) are satisfied for both position
and flux (see \citep[equation~(8)]{men13} for $x_c$ and
equation~(\ref{derilike}) for $\tilde{F}$).  Then we can state the
following result:

\begin{proposition}\label{pro_fish_terms}
The Fisher matrix coefficients for the joint estimation of astrometry
and photometry for a Gaussian PSF, can be written, in exact form, as
follow:
\begin{eqnarray}
\mathcal{I}_{1,1} & = & \frac{1}{2 \pi \sigma^2} \cdot \frac{G F^2}{B} \cdot
\displaystyle \sum_{i=1}^{n} \frac{\left( e^{-\gamma(x^-_i-x_c)} -
  e^{-\gamma(x^+_i-x_c)} \right) ^2}{\left( 1 + \frac{1}{\sqrt{2 \pi}
    \, \sigma}\cdot \frac{F}{B} \cdot J_i(x_c) \right) } \nonumber \\
\mathcal{I}_{1,2} = \mathcal{I}_{2,1} & = & \frac{1}{2 \pi \sigma^2}
\cdot \frac{F}{B} \cdot \displaystyle \sum_{i=1}^{n} \frac{\left(
  e^{-\gamma(x^-_i -x_c)} - e^{-\gamma(x^+_i-x_c)} \right ) \cdot
  J_i(x_c)}{\left( 1 + \frac{1}{\sqrt{2 \pi} \, \sigma}\cdot
  \frac{F}{B} \cdot J_i(x_c) \right) } \nonumber \\
\mathcal{I}_{2,2} & = & \frac{1}{2 \pi \sigma^2} \cdot \frac{1}{GB}
\cdot \displaystyle \sum_{i=1}^{n} \frac{J_i(x_c)^2}{\left( 1 +
  \frac{1}{\sqrt{2 \pi} \, \sigma}\cdot \frac{F}{B} \cdot J_i(x_c)
  \right)} \label{exact}
\end{eqnarray}
(The derivation is presented in Appendix \ref{proof_pro_2D}).
\end{proposition}

In the expressions in equation~(\ref{exact}), we have introduced the
(inverse-)gain of the detector $G$ in units of e$^-$/ADUs (Analog to
Digital Units, or `counts'' on the detector), such that $B$ and $F$
(no tilde) are in ADUs and are defined by $\tilde{F} = G \cdot F$ and
$\tilde{B}= G \cdot B$ respectively. The \crra\ limit in flux,
computed from the above expressions will still be in units of e$^-$.

Note that in the 1D astrometric case, the only meaningful term is
$\mathcal{I}_{1,1}$, which is exactly the inverse of the
\crra\ variance derived in Proposition \ref{pro_astro_1D},
equation~(\ref{eq_sub_sec_astrot1d_2}). Likewise, in the 1D
photometric case, the only meaningful term is $\mathcal{I}_{2,2}$,
which is exactly the inverse of the \crra\ variance as shown by
equation~(\ref{eq_phot1d_2}) in Proposition \ref{pro_photo_1D} above.

\subsection{Analysis and interpretation of the 2D \crra\ bound} \label{subsec_Fisher_int}

In \citet{men13}, it was shown that astrometry is optimal (in the
sense that the positional error budget is minimal), when the object
image is sitting near the edge of a pixel, since positional
information is residing in the slopes of the object image
profile. Interestingly, using the above expressions, we find an effect
for photometry which is just the opposite of that in astrometry: The
lowest variance is found when the source is located towards the center
of a pixel, rather than towards its boundary, this is shown in
Figure~\ref{figpd}. The effect is however quite subtle, and tends to
be worse for severely undersampled images. This could be a relevant
aspect for studies requiring extremely high-accuracy (relative)
photometry (e.g., for observations of exo-planet occultations), and
specially when observing with somewhat undersampled imagers (see
Section~\ref{range}).

At this point it is timely to introduce the definition of
signal-to-noise ratio, $S/N$, as a relevant parameter to interpret the
\crra\ bound. It is possible to show that the $S/N$ for a Gaussian
source is given by\footnote{See \citet[Section 4]{men13}) for details.}:
\begin{equation}
\frac{S}{N} (u_+) = \frac{ P(u_+) \cdot F}{\sqrt{\frac{P(u_+) \cdot
      F}{G} + \frac{u_+}{\sqrt{\ln 2} \, G} \frac{FWHM}{\Delta x}
    \left( f_s \Delta x + \frac{RON^2}{G} \right) }}, \label{sn}
\end{equation}
where $RON$ is the read-out noise per pixel of the detector, in units
of e$^-$, $f_s$ is the sky background (in units of ADUs/arcsec), and
$u_+$ is a dimensionless quantity related to the number of pixels of
the region under which the signal of the source is being measured,
$N_{\mbox{pix}}$, given by (see \citet[equation~(27)]{men13}):

\begin{equation}
u_+ = \frac{1}{\sqrt{\ln 2} \, N_{\mbox{pix}} } \cdot \frac{FWHM}{\Delta
  x}, \label{up}
\end{equation}
and where $P(u_+)$ represents the fraction of the total flux $F$
sampled in the $N_{\mbox{pix}}$, given by $P(u)=\frac{2}{\sqrt{\pi}}
\int_{0}^{u} e^{-v^2} \, dv$.

The overall trend of the 2D \crra\ limit on astrometry and photometry
for one particular choice of parameters is depicted as a function of
the $S/N$ of the source (measured at 90\% of its flux)
in Figure~\ref{figcrap}. As shown in \citet[equation~(45)]{men13}, the
astrometric uncertainty will be either $\propto B^{1/2}/F$ at small
flux (and small $S/N$), or $\propto F^{-1/2}$ at high flux (and large
$S/N$). Therefore, considering the definition of $S/N$, we will have
that $\sigma_{{x}_c} \propto (S/N)^{-1}$. However, as also shown in
the 1D-astrometric setting in \citet[equation~(21)]{men13},
the astrometric \crra\ depends not only on the $S/N$ but also on the
value of the background itself. This is clearly seen on
Figure~\ref{figcrap}, were we compute the \crra\ bound for two values
of $f_s$: We find that, for this choice of parameters, the astrometric
gain by completely suppressing the sky-background (of course an ideal
situation representing the most extreme case one could think of, yet
useful to define strict lower bounds) is significant, almost 20\% in
$\sigma_{{x}_c}$ for both values of the $FWHM$, at a $S/N\sim50$.  The
figure also shows that as the $S/N$ increases the solid and dashed
lines converge, implying that, as the relative importance of the
sky-background becomes smaller, the potential gain in astrometric
accuracy through minimizing the background is reduced, as intuitively
expected.

Rather than looking at the \crra\ limit in flux directly, it is
customary to express this quantity in terms of the uncertainty in
magnitudes, computed as:
\begin{equation}\label{eq_subsec_Fisher_int_1}
 \sigma_{\hat{m}} \equiv \frac{2.5}{2} \cdot
  \left( \log \left( \tilde{F}+\sigma_{{\tilde{F}}} \right) -\log
    \left( \tilde{F}-\sigma_{{\tilde{F}}} \right) \right),
 \end{equation}   
which is quite close to, but in our opinion more robust, than the
classical $\frac{1}{0.4 \ln 10} \cdot
\frac{\sigma_{{\tilde{F}}}}{\tilde{F}}$, since the uncertainties are
not necessarily very small for this last expression to be true. The
results for $\sigma_{\hat{m}}$ are almost indistinguishable from each
other in terms of $FWHM$ or $f_s$, for the choice of parameters in
Figure~\ref{figcrap}. This result is however expected:
Equation~(\ref{fluxcrapp}) shows that the uncertainty in flux will be
either dominated by the square root of the background at small flux
(and small $S/N$), or by the square root of the total flux at high
flux (and large $S/N$). Therefore, since $\sigma_{\hat{m}} \propto
\frac{\sigma_{{\tilde{F}}}}{\tilde{F}}$, we will have that
$\sigma_{\hat{m}} \propto (S/N)^{-1}$, mostly independent of the
background (unlike the case of astrometry, see previous paragraph) or
other parameters. We also note that, after a rapid decline in error as
the $S/N$ increases, the asymptotic behavior of $\sigma_{\hat{m}}$ for
very large $S/N$ may explain, in part, why it is so difficult to
achieve photometric precisions smaller than a few milli-mag. While at
$S/N \sim 100$ we predict $\sigma_{\hat{m}} \sim 11$~[mmag], at $S/N
\sim 200$ we would have $\sigma_{\hat{m}} \sim 5$~[mmag], consistent
with actual measurements, as quoted by \citet{zhet05} (see also
\citet[Section~4 and Table 4.1]{wa06}).

While the expressions for the inverse of the Fisher matrix can be
readily calculated from equation~({\ref{exact}), they do not offer
  much insight into the approximate dependency of the \crra\ bound on
  relevant quantities, like the $FWHM$ or the $S/N$ of the source, or
  the detector pixel size $\Delta x$. For this purpose, it is useful
  to resort to the small pixel (high resolution) approximation of a
  Gaussian PSF, which is done in the next Section.

\subsection{The 2D \crra\ bound in the small pixel (high resolution)
  approximation} \label{crhr}

If we assume that the pixel array oversamples the source, i.e., if
$\Delta x / \sigma \ll 1$, then
one has that:
\begin{eqnarray}
g_i(x_c)                              & \simeq & \Phi(x_i-x_c) \cdot \Delta x \\
\frac{\partial g_i(x_c)}{\partial x_c} & \simeq & \frac{(x_i-x_c)}{\sigma^2} \cdot g_i(x_c).
\end{eqnarray}

As it can be easily verified, under this approximation the elements of
the Fisher matrix become:

\begin{eqnarray} \label{fishhr}
\mathcal{I}_{1,1} & = & \frac{\tilde{F}^2}{\sigma^4} \cdot \displaystyle \sum_{i=1}^n
\frac{(x_i-x_c)^2}{( \tilde{F} g_i(x_c) + \tilde{B}_i )} \cdot g_i(x_c)^2 \nonumber\\
\mathcal{I}_{1,2} & = & \frac{\tilde{F}}{\sigma^2} \cdot \displaystyle
\sum_{i=1}^n \frac{(x_i-x_c)}{( \tilde{F} g_i(x_c) + \tilde{B}_i )} \cdot
g_i(x_c)^2 \nonumber \\
\mathcal{I}_{2,2} & = & \displaystyle \sum_{i=1}^n \frac{1}{( \tilde{F} g_i(x_c) +
  \tilde{B}_i )} \cdot g_i(x_c)^2 .
\end{eqnarray}

We note that the term $\mathcal{I}_{1,2}$ in equation~(\ref{fishhr}),
being a function of an odd power of $(x_i-x_c)$, is expected to be
very small if the source is well sampled by the detector, an important
fact that will be fully exploited in the following analysis (see also
Section~\ref{range}).
On the other hand, the dependence on $\tilde{F} g_i(x_c) +
\tilde{B}_i$ in the denominator of equation~(\ref{fishhr}) makes it
difficult to get simple analytical expressions for them. However,
things simplify notably in the two extreme regimes of flux- and
background-dominated sources, which we will examine in turn in the
next sub-sections.

\subsubsection{Flux dominated sources in the Small Pixel (High
  Resolution) approximation}
\label{subsub_flux}  

In this case, a first-order series development of the term $(\tilde{F}
g_i(x_c) + \tilde{B}_i)^{-1}$ in equation~(\ref{fishhr}), in terms of
the quantity $\tilde{B}/\tilde{F}$ (assumed to be $\ll 1$), yields the
following:
%
%
\begin{eqnarray} \label{fishhrf1}
\mathcal{I}_{1,1} & = & \frac{\tilde{F}}{\sigma^4} \cdot \left( \displaystyle
\sum_{i=1}^n (x_i-x_c)^2 \cdot g_i(x_c) - \frac{\tilde{B}}{\tilde{F}} \cdot
  \displaystyle \sum_{i=1}^n (x_i-x_c)^2  \right) \nonumber \\
\mathcal{I}_{1,2} & = & \frac{1}{\sigma^2} \cdot \left( \displaystyle
\sum_{i=1}^n (x_i-x_c) \cdot g_i(x_c) - \frac{\tilde{B}}{\tilde{F}} \cdot
  \displaystyle \sum_{i=1}^n (x_i-x_c)  \right) \nonumber \\
\mathcal{I}_{2,2} & = & \frac{1}{\tilde{F}} \cdot \left(1 - n \cdot
\frac{\tilde{B}}{\tilde{F}}  \right) .
\end{eqnarray}

This series development allows us to write some of the terms in the
above expressions in an analytical closed-form, which greatly
facilitates the evaluation of the \crra\ bound. For our Gaussian PSF
we will have, in the high resolution approximation, that:
\begin{eqnarray} \label{sumx}
\displaystyle \sum_{i=1}^n (x_i-x_c) \cdot g_i(x_c) & \approx & \frac{1}{\sqrt{2
    \pi} \sigma} \cdot \lim_{\Delta x \rightarrow 0} \sum_{i=1}^n (x_i-x_c)
\cdot e^{-\frac{(x_i-x_c)^2}{2 \, \sigma^2}} \cdot \Delta x \nonumber \\
& = & \frac{1}{\sqrt{2 \pi} \sigma} \cdot \int_{-\infty}^{+\infty}
(x - x_c) \cdot e^{-\frac{(x-x_c)^2}{2 \, \sigma^2}} \, dx \nonumber \\
& = & 0,
\end{eqnarray}
while, on the other hand:
\begin{eqnarray} \label{sumx2}
\displaystyle \sum_{i=1}^n (x_i-x_c)^2 \cdot g_i(x_c) & \approx & \frac{1}{\sqrt{2
    \pi} \sigma} \cdot \lim_{\Delta x \rightarrow 0} \sum_{i=1}^n (x_i-x_c)^2
\cdot e^{-\frac{(x_i-x_c)^2}{2 \, \sigma^2}} \cdot \Delta x \nonumber \\
& = & \frac{1}{\sqrt{2 \pi} \sigma} \cdot \int_{-\infty}^{+\infty}
( x - x_c)^2 \cdot e^{-\frac{(x-x_c)^2}{2 \, \sigma^2}} \, dx \nonumber \\
& = & \sigma^2 .
\end{eqnarray}
Replacing (\ref{sumx}) and~(\ref{sumx2}) into
equation~(\ref{fishhrf1}), we end up with:
\begin{eqnarray} \label{fishhrf2}
\mathcal{I}_{1,1} & = & \frac{\tilde{F}}{\sigma^4} \cdot \left( \sigma^2 -
\frac{\tilde{B}}{\tilde{F}} \cdot \displaystyle \sum_{i=1}^n
(x_i-x_c)^2 \right) \nonumber \\
\mathcal{I}_{1,2} & = & - \frac{1}{\sigma^2} \cdot \frac{\tilde{B}}{\tilde{F}}
  \cdot \displaystyle \sum_{i=1}^n (x_i-x_c) \nonumber \\
\mathcal{I}_{2,2} & = & \frac{1}{\tilde{F}} \cdot \left(1 - n \cdot
\frac{\tilde{B}}{\tilde{F}}  \right).
\end{eqnarray}

With these coefficients, it is easy to see that the determinant,
required for the evaluation of the inverse of the Fisher matrix, can
be written, to first order in $\tilde{B}/\tilde{F}$, as follows:
\begin{equation}
\mathcal{I}_{11} \cdot \mathcal{I}_{22} - \mathcal{I}_{12}^2 \approx
\frac{1}{\sigma^4} \cdot \left( \sigma^2 - \frac{\tilde{B}}{\tilde{F}}
\cdot \displaystyle \sum_{i=1}^n (x_i-x_c)^2 \right) \cdot \left( 1 -
n \cdot \frac{\tilde{B}}{\tilde{F}} \right),
\end{equation}
then, it can be verified that the \crra\ bound for astrometry becomes:
\begin{equation} \label{crxf}
\sigma^2_{{x}_c} = [ \mathcal{I}_{(x_c,\tilde{F})}(n)^{-1} ]_{1,1}
\approx \frac{1}{8 \ln 2} \cdot \frac{1}{G F} \cdot \left(1 +
\frac{1}{8 \ln 2} \cdot \frac{B}{F} \cdot \frac{ \sum_{i=1}^n
  (x_i-x_c)^2}{FWHM^2} \right) \cdot FWHM^2.
\end{equation}

We note that equation~(\ref{crxf}) is equivalent to
equation~(\ref{eq_sub_sec_astrot1d_3}-bottom line) for the 1D case,
but where the extra term in parenthesis in equation~(\ref{crxf})
accounts for the fact that in the present case we have retained the
terms up to first order in the small quantity (i.e., if $B/F
\rightarrow 0$, both equations coincide exactly). The validity of
equation~(\ref{crxf}) in comparison with both prior theoretical
estimates as well as real astrometry, has already been discussed in
\citet{men13}. Here we would like to add that the results by
\citet{gaet85}, based on measurements with the Multichannel
Astrometric Photometer, fully support our prediction that the
astrometric accuracy improves as the inverse of the square root of the
photon counts, as predicted by equation~(\ref{crxf}).

Completely analogously, the \crra\ bound for photometry, provided that
$n \cdot B \ll F$ (meaning that the flux is being estimated within a
reasonable aperture, containing most of the flux, but avoiding to
incorporate background far away from the main core of the source),
becomes:
\begin{equation} \label{crff}
\sigma^2_{\tilde{F}} = [ \mathcal{I}_{(x_c,\tilde{F})}(n)^{-1} ]_{2,2}
\approx G F \cdot \left( 1 + n \cdot \frac{B}{F} \right).
\end{equation}

This equation is equivalent to the 1D expression shown in
equation~(\ref{fluxcrapp}-bottom line), and it shows that, in the
small pixel approximation, the determination of the flux is completely
decoupled from the astrometry (see Section~\ref{range} for a further
discussion of this), leading to the well-known fact that the expected
standard deviation of the flux goes as the square root of the flux
itself (measured in e$^{-}$) when the source dominates the counts, a
feature which is characteristic of a Poisson-driven detection process.

\subsubsection{Background dominated sources in the Small Pixel (High
  Resolution) approximation} \label{bdsp}

Let us know explore the other regime, i.e., when $\tilde{F}/\tilde{B}
\ll 1$. Following the same steps as in the previous section, it is
simple to verify that equation~(\ref{fishhr}) become:
\begin{eqnarray} \label{fishhrb1}
\mathcal{I}_{1,1}  & = & \frac{\tilde{F^2}}{\tilde{B} \sigma^4} \cdot \left( \displaystyle
\sum_{i=1}^n (x_i-x_c)^2 \cdot g_i(x_c)^2 - \frac{\tilde{F}}{\tilde{B}} \cdot
  \displaystyle \sum_{i=1}^n (x_i-x_c)^2 \cdot g_i(x_c)^3 \right) \nonumber \\
\mathcal{I}_{1,2}  & = & \frac{\tilde{F}}{\tilde{B} \sigma^2} \cdot \left( \displaystyle
\sum_{i=1}^n (x_i-x_c) \cdot g_i(x_c)^2 - \frac{\tilde{F}}{\tilde{B}} \cdot
  \displaystyle \sum_{i=1}^n (x_i-x_c) \cdot g_i(x_c)^3  \right) \nonumber \\
\mathcal{I}_{2,2}  & = & \frac{1}{\tilde{B}} \cdot \left( \displaystyle
\sum_{i=1}^n g_i(x_c)^2 - \frac{\tilde{F}}{\tilde{B}} \cdot \displaystyle
\sum_{i=1}^n g_i(x_c)^3 \right).
\end{eqnarray}

It can be readily seen from equation~(\ref{sumx}) that a summation
involving any power of $g_i(x_c)$, modulated by an odd function of
$(x_i-x_c)$ will be zero in the high resolution regime, and therefore
the off-diagonal term $\mathcal{I}_{1,2}$ in equation~(\ref{fishhrb1})
will be zero in this case. The other summations in
equation~(\ref{fishhrb1}) can be easily calculated following the same
procedure outlined in
(\ref{sumx}) and~(\ref{sumx2}), for example:
\begin{eqnarray}
\displaystyle \sum_{i=1}^n g_i(x_c)^2 & \approx & \frac{\Delta x}{2 \pi \sigma^2}
\cdot \lim_{\Delta x \rightarrow 0} \sum_{i=1}^n
e^{-\frac{(x_i-x_c)^2}{\sigma^2}} \cdot \Delta x \nonumber \\
& = & \frac{\Delta x}{2 \pi \sigma^2} \cdot \int_{-\infty}^{+\infty}
e^{-\frac{(x-x_c)^2}{\sigma^2}} \, dx \nonumber \\
& = & \frac{1}{2 \sqrt{\pi}} \cdot \frac{\Delta x}{\sigma},
\end{eqnarray}
and, also:
\begin{eqnarray}
\displaystyle \sum_{i=1}^n (x_i-x_c)^2 \cdot g_i(x_c)^2 & \approx & \frac{\Delta
  x}{2 \pi \sigma^2} \cdot \lim_{\Delta x \rightarrow 0} \sum_{i=1}^n
(x_i-x_c)^2 \cdot e^{-\frac{(x_i-x_c)^2}{\sigma^2}} \cdot \Delta x
\nonumber \\
& = & \frac{\Delta x}{2 \pi \sigma^2} \cdot
\int_{-\infty}^{+\infty} ( x - x_c)^2 \cdot
e^{-\frac{(x-x_c)^2}{\sigma^2}} \, dx \nonumber \\
& = & \frac{1}{4 \sqrt{\pi}} \cdot \Delta x \cdot \sigma.
\end{eqnarray}

The other terms in equation~(\ref{fishhrb1}) can be calculated in an
analogous way, obtaining:
\begin{eqnarray}
\displaystyle \sum_{i=1}^n g_i(x_c)^3 & \approx & \frac{1}{2 \sqrt{3} \pi} \cdot
\left( \frac{\Delta x}{\sigma} \right)^2 , \\
\displaystyle \sum_{i=1}^n (x_i-x_c)^2 \cdot g_i(x_c)^3 & \approx &
\frac{\Delta x^2}{6 \sqrt{3} \pi}.
\end{eqnarray}

Using the above values for the coefficients, the \crra\ bound for
astrometry is given by:
\begin{equation} \label{crxb}
\sigma^2_{{x}_c} = [ \mathcal{I}_{(x_c,\tilde{F})}(n)^{-1}
]_{1,1} \approx \frac{1}{4 \ln 2} \sqrt{ \frac{\pi}{2 \ln 2} } \cdot \frac{B}{G F^2}
\cdot \left( 1 + \frac{4}{3} \sqrt{ \frac{2 \ln 2}{3 \pi} } \cdot
\frac{F}{B} \cdot \frac{\Delta x}{FWHM} \right) \cdot
\frac{FWHM^3}{\Delta x}.
\end{equation}

As it was mentioned in Section \ref{subsub_flux}, here too, this
equation reduces to the 1D equation in
(\ref{eq_sub_sec_astrot1d_3}-top line), when the ratio $F/B
\rightarrow 0$.  On the other hand, the corresponding \crra\ bound for
photometry would be, in this case:
\begin{equation} \label{crfb}
\sigma^2_{\tilde{F}} = [ \mathcal{I}_{(x_c,\tilde{F})}(n)^{-1}
]_{2,2} \approx \sqrt{\frac{\pi}{2 \ln2} } \cdot GB \cdot \left( 1 + 2
\sqrt{\frac{2 \ln2}{3 \pi}} \cdot \frac{F}{B} \cdot \frac{\Delta
  x}{FWHM} \right) \cdot \frac{FWHM}{\Delta x}.
\end{equation}

In equation~(\ref{crfb}), the ratio $(FWHM/\sqrt{\ln 2} \, \Delta x)$
represents the sampling of the PSF of the object\footnote{More details
  in \citet[equation~(27)]{men13}.} and we can see, from
equation~(\ref{up}) that $u_+ = \sqrt{\pi/2} \approx 1.253$, which
represents an aperture containing $\sim 92$\% of the equivalent
``flux'' (given by $P(u_+)$). So, in this setting, too (as it was the
case of equation~(\ref{crff})), the uncertainty in the flux goes as
square root of the flux, which is however in this case mostly provided
by the background. We also note that the term $B/\Delta x$ is
approximately equal to the sky background in units of ADU/arcsec,
therefore equation~(\ref{crfb}) implies a total aperture (diameter)
that samples $\sqrt{\frac{\pi}{2 \ln2} } \approx 1.5$ times the $FWHM$
of the source.

\subsection{Range of use of the high resolution \crra\ bound} \label{range}

Given the simplicity of the equations derived in the previous
sub-sections, it is important to define how quickly
equations~(\ref{crxf}) and~(\ref{crff}), or~(\ref{crxb})
and~(\ref{crfb}) deteriorate as we move away from their respective
regimes of application. Interestingly enough, the approximate flux
\crra\ bound is a lot more insensitive to the assumptions involved
than its astrometric counterpart. For example, a very weak source
($S/N=3$) with a $FWHM=1.0$~arcsec has a predicted \crra\ bound
uncertainty in flux of $\sim$27\%, the same value as derived from
equation~(\ref{crfb}), independently of $\Delta x$ from 0.1 to
1.0~arcsec. In the same regime, the astrometric \crra\ bound increases
from $\sigma_{{x}_{c}}=164$~mas to 321~mas for $\Delta x$ from 0.1 to
1.0~arcsec, whereas equation~(\ref{crxb}) predicts
$\sigma_{{x}_{c}}=160$~mas (1 mas = 1 milli-arcsec). For strong
sources ($S/N=200$) the ``exact'' photometric \crra\ bound (derived
from equation~(\ref{exact})) predicts 0.5\% uncertainty in flux, same
as that given by equation~(\ref{crff}), whereas the exact astrometric
\crra\ calculation shows an increase from 2.2 to 3.1~mas when $\Delta
x$ increases from 0.1 to 1.0~arcsec, while that predicted by
equation~(\ref{crxf}) gives 1.9~mas. As a rule of thumb we find that,
as long as $\Delta x / FWHM < 0.5$, equations~(\ref{crxf})
and~(\ref{crff}), or~(\ref{crxb}) and~(\ref{crfb}) are quite reliable,
and can be very useful as quick estimators.

Another aspect of the above discussion is that, as can be seen from
the analysis presented in the previous two sub-Sections, a critical
assumption of the adopted approximations in the high-resolution regime
is the fact that, to first order on either $\tilde{B}/\tilde{F}$ or
$\tilde{F}/\tilde{B}$, the coefficient $\mathcal{I}_{1,2} \sim
0$. Since the off-diagonal terms in the Fisher matrix represent the
strength of the co-dependency among the parameters to be estimated (in
this case position and flux), in practice this means that, under this
assumption, the \crra\ bound in astrometry and photometry become
de-coupled from each other, thus converging to their respective 1D
approximations. It is therefore interesting to explore approximately
under which regime of parameters this assumption actually holds. For
this purpose, in Figure~\ref{figdiffa} we show the behavior of the
difference between the exact 2D expressions derived in
Section~\ref{main_section}, denoted by $\sigma_{{x}_{c}}$ and
$\sigma_{{\tilde{F}}}$, and their exact 1D, counterparts
(equations~(\ref{eq_phot1d_2}) and (\ref{eq_sub_sec_astrot1d_2})),
denoted by $\sigma_{{x}_{c_{1D}}}$ and $\sigma_{{\tilde{F}}_{1D}}$,
computed for astrometry and photometry respectively as:
\begin{eqnarray}
\Delta \sigma_x \equiv
\frac{\sigma_{{x}_c}-\sigma_{{x}_{c_{1D}}}}{\sigma_{{x}_c}} & = &
\frac{ \sqrt{\frac{\mathcal{I}_{2,2}}{\mathcal{I}_{1,1} \cdot
      \mathcal{I}_{2,2} - \mathcal{I}_{1,2}^2}} \, - \,
  \sqrt{\frac{1}{\mathcal{I}_{1,1}}}}{\sigma_{{x}_c}}, \nonumber \\ \Delta
\sigma_{\tilde{F}} \equiv
\frac{\sigma_{{\tilde{F}}}-\sigma_{{\tilde{F}}_{1D}}}{\sigma_{{\tilde{F}}}}
& = & \frac{\sqrt{\frac{\mathcal{I}_{1,1}}{\mathcal{I}_{1,1} \cdot
      \mathcal{I}_{2,2} - \mathcal{I}_{1,2}^2}} \, - \,
  \sqrt{\frac{1}{\mathcal{I}_{2,2}}}}{\sigma_{{\tilde{F}}}}.
\end{eqnarray}

Note that, defined this way, both are dimensionless fractional
quantities, and should be always $\ge 0$.

It turns out that, numerically, the fractional values for $\Delta
\sigma_x$ and $\Delta \sigma_{\tilde{F}}$ are quite similar as a
function of $\Delta x$. As an example, in Figure~\ref{figdiffa}, we
see that the effect of neglecting the cross-term $\mathcal{I}_{1,2}$
for this particular setting, having a relatively high $S/N$, is only
noticeable for under-sampled images but, as can be seen from the
figure, in any case the difference is smaller than $\sim$15\% under a
wide-variety of reasonable conditions (see below for further details
on this). We also notice that, as expected, the differences $\Delta
\sigma_{x}$ and $\Delta \sigma_{\tilde{F}}$ depend (in a complex way)
on the pixel offset, illustrating the effect of symmetry breaking in
odd terms involving $(x_i-x_c)$ (recall Section~\ref{bdsp}). We have
verified that, at lower $S/N$ ($\sim 6$), the effect is much steeper,
and it occurs at smaller $\Delta x$, but it is still true that, for
well sampled images ($\Delta x < FWHM$), the differences are minimal
(less than 1\%). The overall corollary of this exercise is that the 1D
\crra\ case for photometry and astrometry can be safely used for quick
estimation purposes, instead of their more complex 2D cousin, being
quite forgiving about the fine-pixel requirement $\Delta x / \sigma
\ll 1$.

Since some recent large-area surveys and robotic telescopes are
exploring the undersampled regime, e.g.,
SuperWASP\footnote{http://www.superwasp.org/}, 13.7~arcsec/pix
(described by \citet{poll06}),
TRAPPIST\footnote{http://www.orca.ulg.ac.be/TRAPPIST/},
0.64~arcsec/pix (described by \citet{gill11}), the Catalina Real-Time
Transient Survey\footnote{http://crts.caltech.edu/}, 0.98, 1.84 and
2.57 arcsec/pix (described by \citet{djor11}, or the La Silla-QUEST
Variability Survey\footnote{http://hep.yale.edu/lasillaquest},
0.88~arcsec/pix (described by \citet{balt13}) among others, it is
interesting to quantify the impact of this design feature into the
predicted \crra\ bound.  To estimate the effect of neglecting the
cross-dependency between flux and astrometry, Table~\ref{tabunder}
compares the 1D and 2D \crra\ limits as a function of the pixel size
$\Delta x$, and the $S/N$ of the source, adopting the same parameters
as those of Figure~\ref{figdiffa}.
In the table, the astrometric \crra\ is in units of mas, whereas the
\crra\ bound in flux is in \%, defined by $100 \cdot
\frac{\sigma_{{\tilde{F}}}}{\tilde{F}}$ and $100 \cdot
\frac{\sigma_{{\tilde{F}_{1D}}}}{\tilde{F}_{1D}}$ respectively. Since,
as discussed previously (see also Figure~\ref{figdiffa}), the
\crra\ limit depends on the centering of the source on the pixel, we
have computed the \crra\ limit for two representative pixel offsets,
of 0.125~pix and 0.25~pix. As it can be seen from this table, at
intermediate and high $S/N$ the photometry is not appreciably affected
by the pixel size, but we naturally see a gradual deterioration of the
location accuracy as the pixel size increases. At low $S/N$ the impact
of pixel size (and pixel offsets) becomes critical for astrometry, and
noticeable for photometry. Across the table we also see the impact of
pixel offsets on the expected precision for both photometry and
astrometry, in particular a ``feature'' already discussed in this and
in our previous paper, namely that astrometry is better done near the
pixel boundaries (large offsets), whereas photometry is better done
near the pixel centers (small offsets). At a low $S/N=5$ and very
undersampled images ($\Delta x = 1.5$~arcsec), one may even argue that
pixels offsets can make the difference between non-detection and
detection of the source: Compare the formal astrometric \crra\ value
for a pixel offset of 0.125~pix, with $\sigma_{x_c} \sim 4$~arcsec, to
the more reasonable value of $\sigma_{x_c} \sim 0.5$~arcsec for a
pixel offset of 0.25~pix.

\subsection{Effects of a variable PSF or a variable background} \label{variable}

So far we have assumed that the PSF, mostly characterized in our
scheme by its $FWHM$, is constant across the detector. However, in
many cases, the telescope plus camera optical system may introduce
variations in the $FWHM$ of the images at the focal plane
\citep{sch13}, and even changes on the shape of the PSF (e.g.,
aberrations). Also, focal reducers, commonly used in wide-field
imagers, can introduce illumination problems that generate background
variations on scales of the field-of-view of the detector
\citep{sel04}. Both of these effects will have an impact on the
\crra\ bound, depending on the position of the source relative to the
optical axis of the camera, and it is therefore important to quantify
them.

In the case of oversampled images, the effect on astrometry and
photometry of changes in the $FWHM$ and the (local) background $B$ can
be readily calculated from equations~(\ref{crxf}) and~(\ref{crff}),
or~(\ref{crxb}) and~(\ref{crfb}). From these we see that, at high
$S/N$, the astrometric \crra\ bound scales approximately linearly with
the $FWHM$ (while the photometry is independent of the $FWHM$),
whereas at low $S/N$ the impact on the expected astrometric precision
due to changes on the width of the PSF gets amplified by a factor of
1.5. On the other hand, for well-exposed images, small background
variations do not have an important impact on astrometry nor
photometry, as intuitively expected, whereas for weak images we have
that $\frac{\Delta \sigma_{x_c}}{\sigma_{x_c}} = \frac{\Delta
      \sigma_{\tilde{F}}}{\tilde{F}} = \frac{1}{2} \cdot \frac{\Delta
    B}{B}$.

For undersampled images, we have to resort to the exact expressions,
given by equation~(\ref{exact}). In Table~(\ref{tabpsfbac}) we show
the effect of a change of 20\% in the width of the PSF, or a 10\%
change in the background, on the predicted photometric and astrometric
\crra\ bounds, for an under-sampled image, with $FWHM=0.5$ and $\Delta
x = 1.0$~arcsec. We have computed this for the best-case scenario for
photometry (source centered on a given pixel, upper part of the
table), and for the best case scenario for astrometry (source centered
on a pixel boundary, lower part of the table). As it can be seen from
the table, the impact of these changes on the photometry (provided
that the background is properly accounted for in the photometric
measurements), is minimal. On the other hand, for the best case
astrometric setting, the 10\% change in the $FWHM$ implies a
$\sim20$\% change in the astrometric \crra\ limit, whereas this
increases to as much as 50\% for the worst case centering. Changes in
background have a smaller, albeit non-negligible, impact on the
astrometry, inducing a 5\% increase in the \crra\ limit regardless of
the centering location. These results are at variance with the
high-resolution behavior (see previous paragraph), which shows the
importance of computing the \crra\ bound in this specific situation
for each particular case.

We finally note that, in all calculations above, we have still assumed
a Gaussian PSF. A meaningful extension to other PSF shapes requires an
extension of the \crra\ calculation to a fully two-dimensional X-Y
array, including the possibility of a cross-correlation term in the
PSF between the X and Y coordinates (i.e., that the shape of the PSF
is not necessarily oriented along any of the CCD axis, case of
aberrated images), which we hope to explore in forthcoming papers.

\section{Conclusions} \label{conc}

We have developed general expressions for the \crra\ minimum variance
bound for the joint estimation of photometry and astrometry in a
linear detector for a Gaussian source.

We show that the minimum expected photometric errors depend on the
position of the source with respect to the pixel center, being larger
if the source is located toward the pixel boundaries. The effect is
subtle, and becomes more relevant for undersampled images. This result
is exactly the opposite of what is found for the astrometric
\crra\ bound, and described thoroughly in \citet[Section 3.3]{men13}.

We demonstrate that both, astrometric and photometric (magnitudes)
minimal error bounds, vary $\propto (S/N)^{-1}$, while the astrometry
is, additionally, quite sensitive to the value of the background -
suppressing the background can greatly enhance the astrometric
accuracy.

When the detector adequately samples the source (oversampling
regime), we show that the joint parametric determination of
photometry and astrometry for the source become decoupled from each
other, and furthermore, it is possible to write down closed-form
expressions (approximate to first order in the small quantities $F/B$
or $B/F$) for the expected minimum uncertainty estimation of the flux and
position. We formally verify the known fact that the uncertainty in flux
depends {\it mostly} on the square root of the flux, while for
astrometry we recover the astrometry-only 1D \crra\ results found by
\citet{men13}.

We show that the de-coupling of the \crra\ bounds between
$\sigma_{{x}_c}$ and $\sigma_{\tilde{F}}$ is quite resilient to the
assumption $\Delta x / FWHM \ll 1$ and, in fact, as long as we satisfy
$\Delta x / FWHM < 0.5$, the cross term $\mathcal{I}_{1,2}$ in
the Fisher information matrix is negligible. Given this result, we
regard equations~(\ref{crxf}), (\ref{crff}), (\ref{crxb}) and
(\ref{crfb}) as particularly useful benchmark estimators for the
maximum attainable photometric and astrometric precision, given a
detector setting and pre-specified observational conditions.

Finally, we explore the impact of variations in the spread of the PSF,
or on the level of the background, upon the \crra\ limit, and we
derive expressions for the precision bound in some simple cases. We
also show that, in general, astrometry is more sensitive
(fractionally) than photometry due to variations in the $FWHM$ or the
background.

\newpage

\appendix
\section{Proof of Proposition \ref{pro_photo_1D}: \crra\ bound for flux}
\label{proof_pro_photo_1D}

In order to insure that the conditions for the \crra\ bound are met,
we need to verify that the constraint in (\ref{cond2d}) is satisfied
in regards to the parameter $\tilde{F}$ and, if so, we are allowed to
use equations~(\ref{varcr}) and~(\ref{fisher}) to compute the
\crra\ bound. Using equation~(\ref{eq_sub_sec_joint_estimation_1}) and
omitting the explicit dependency on $x_c$ on all the variables there,
we have for the parameter $\tilde{F}$ that:
\begin{eqnarray}
\frac{d \ln L(\vec{I};\tilde{F}) }{d \tilde{F}} & = & \frac{d}{d
  \tilde{F}} \left( {\sum_{i=1}^n \left( I_i\cdot \ln
  \lambda_i(\tilde{F}) - \lambda_i(\tilde{F}) - \ln I_i! \right) }
\right) \label{derilike1} \\ & = & \sum_{i=1}^n I_i \cdot
\frac{1}{\lambda_i(\tilde{F})} \cdot \frac{d \lambda_i(\tilde{F})}{d
  \tilde{F}} - \sum_{i=1}^n \frac{d \lambda_i(\tilde{F})}{d
  \tilde{F}}. \label{derilike}
\end{eqnarray}
If $\mathbb{E}$ is the expected value with respect to the vector of
observables $(I_1,...,I_n)$ given $\tilde{F}$, we indeed verify from
the above expression that $\mathbb{E}_{I_1,...,I_n} \left( \frac{d \ln
  L(\vec{I};\tilde{F})}{d \tilde{F}} \right) =0$ because
$\mathbb{E}({I_i})=\lambda_i(\tilde{F})$.  Hence, we can apply
equations~(\ref{varcr}) and (\ref{fisher}).

First, we need to compute the Fisher information (\ref{fisher}) of the data about
$\tilde{F}$, which is given by: 
\begin{equation} \label{a1}
\mathcal{I}_{\tilde{F}}(n) \equiv
  \mathbb{E}_{I_1,...,I_n\sim f_{\tilde{F}}^n} \left( \left( \frac{d
        \ln L(\vec{I};\tilde{F}) }{d \tilde{F}} \right) ^2 \right)
\end{equation}
Noting that $d \lambda_i(\tilde{F})/d \tilde{F} = g_i$, from
~(\ref{derilike}) we will thus have:
\begin{equation} \label{a2}
\frac{d \ln L(\vec{I};\tilde{F}) }{d \tilde{F}} = \sum_{i=1}^n g_i \cdot
\frac{I_i}{\lambda_i} - 1, 
\end{equation}
where we have used the fact that $\sum_{i=1}^n g_i = 1$ (see
equation~(\ref{eq_sub_sec_joint_estimation_5})). From this we can
write:
\begin{eqnarray} \label{a3}
\left( \frac{d \ln L(\vec{I};\tilde{F}) }{d \tilde{F}} \right) ^2 & =
& \left( \sum_{i=1}^n g_i \cdot \frac{I_i}{\lambda_i} \right)^2 - 2
\cdot \sum_{i=1}^n g_i \cdot \frac{I_i}{\lambda_i} + 1 \nonumber \\
& = &  \sum_{i=1}^n \sum_{j=1}^n g_i g_j \cdot \frac{I_i
  I_j}{\lambda_i \lambda_j}  - 2 \cdot  \sum_{i=1}^n
g_i \cdot \frac{I_i}{\lambda_i}  + 1 \nonumber \\
& = &  \sum_{i=1}^n g_i^2 \cdot \frac{I_i^2}{\lambda_i^2} + 
 \sum_{i=1}^n \sum_{j \neq i}^n g_i g_j \cdot \frac{I_i
  I_j}{\lambda_i \lambda_j}  - 2 \cdot  \sum_{i=1}^n
g_i \cdot \frac{I_i}{\lambda_i}  + 1.
\end{eqnarray}
Therefore,
\begin{eqnarray} \label{a4}
\mathbb{E}\left( \frac{d \ln L(\vec{I};\tilde{F}) }{d \tilde{F}}
\right) ^2 & = & \sum_{i=1}^n g_i^2 + \sum_{i=1}^n g_i^2 \cdot
\frac{1}{\lambda_i} + \sum_{i=1}^n \sum_{j \neq i}^n g_i g_j - 1
\nonumber \\
& = & \left( \sum_{i=1}^n g_i \right)^2 + \sum_{i=1}^n g_i^2 \cdot
\frac{1}{\lambda_i} - 1 \nonumber \\
& = &  \sum_{i=1}^n \frac{g_i^2}{\lambda_i}
\end{eqnarray}
where we have used the facts that $\mathbb{E}(I_i)= \lambda_i$,
$\mathbb{E}(I_i^2)= \lambda_i^2 + \lambda_i$, and $\mathbb{E}(I_i
\cdot I_j) = \mathbb{E}(I_i) \cdot \mathbb{E}(I_j) = \lambda_i \cdot
\lambda_j$, this last expression since the pixel measurements are
independent. With the above expression, and equation~(\ref{a1}), we
see that:
\begin{equation}
  \mathcal{I}_{\tilde{F}}(n) = \displaystyle \sum_{i=1}^{n} \frac{
    g_i^2} { \left( \tilde{F} \cdot g_i + \tilde{B}_{i} \right) },
\end{equation}
from which the expression in equation~(\ref{eq_phot1d_1}), namely
$\sigma^2_{\tilde{F}_{1D}} = \mathcal{I}_{\tilde{F}}(n)^{-1}$, follows
directly.

\newpage

\section{Proof of Proposition \ref{pro_fish_terms}: Fisher information matrix for joint astrometry and flux.}

\label{proof_pro_2D}

The likelihood function is given by $\ln L(\vec{I}; (x_c,\tilde{F})) =
{\sum_{i=1}^n \left( I_i\cdot \ln \lambda_i(x_c,\tilde{F}) -
  \lambda_i(x_c,\tilde{F}) - \ln I_i!  \right) }$. The required
partial derivatives are given by:
\begin{equation}
\frac{\partial \ln L(\vec{I}; (x_c,\tilde{F}))}{\partial x_c} =
\sum_{i=1}^n \left(\frac{I_i}{\lambda_i(x_c,\tilde{F})} \cdot \frac{\partial
  \lambda_i(x_c,\tilde{F})}{\partial x_c} - \frac{\partial
  \lambda_i(x_c,\tilde{F})}{\partial x_c} \right),
\end{equation}
and,
\begin{equation}
\frac{\partial \ln L(\vec{I}; (x_c,\tilde{F}))}{\partial \tilde{F}} =
\sum_{i=1}^n \left( \frac{I_i}{\lambda_i(x_c,\tilde{F})} \cdot g_i(x_c)
- g_i(x_c) \right),
\end{equation}
where we have used the fact that since, by definition,
$\lambda_i(x_c,\tilde{F}) = \tilde{F} \cdot g_i(x_c) + \tilde{B}_i$,
then $\partial \lambda_i(x_c,\tilde{F})/\partial \tilde{F} =
g_i(x_c)$.

We verify that, both $\mathbb{E}_{I_1,...,I_n} \left( \frac{\partial
  \ln L(\vec{I};\tilde{F})}{\partial x_c} \right) =0$ and
$\mathbb{E}_{I_1,...,I_n} \left( \frac{\partial \ln
  L(\vec{I};\tilde{F})}{\partial \tilde{F}} \right) =0$ because
$\mathbb{E}({I_i})=\lambda_i(\tilde{F})$.  Hence, we can apply
equations~(\ref{varcr}) and (\ref{fisher}). To make mathematical notation
easier, in what follows we identify the sub-index '$1$' with the
parameter of spatial coordinate $x_c$, while the sub-index '$2$'
refers to the parameter flux $\tilde{F}$.  Consequently, 
the individual matrix terms are:
\begin{eqnarray}\label{a11}
\mathcal{I}_{1,1}(n) \equiv & \mathbb{E}_{I_1,...,I_n} \left( \left(
\frac{\partial \ln L(\vec{I};x_c,\tilde{F}) }{\partial x_c}
\right)^2 \right) \nonumber\\
=  & \mathbb{E}_{I_1,...,I_n} \left( \sum_i \sum_j
\left( \frac{I_i \cdot I_j}{\lambda_i(x_c,\tilde{F}) \cdot
  \lambda_j(x_c,\tilde{F})} \cdot \frac{\partial
  \lambda_i(x_c,\tilde{F})}{\partial x_c} \cdot \frac{\partial
  \lambda_j(x_c,\tilde{F})}{\partial x_c} - 2 \cdot
\frac{I_i}{\lambda_i(x_c,\tilde{F})} \cdot \frac{\partial
  \lambda_i(x_c,\tilde{F})}{\partial x_c} \cdot \frac{\partial
  \lambda_j(x_c,\tilde{F})}{\partial x_c} \right) \right) \nonumber\\
& + \sum_i\sum_j \frac{\partial \lambda_i(x_c,\tilde{F})}{\partial x_c} \cdot
\frac{\partial \lambda_j(x_c,\tilde{F})}{\partial x_c} \nonumber\\
= & \mathbb{E}_{I_1,...,I_n} \left( \sum_i \left(\frac{I_i
}{\lambda_i(x_c,\tilde{F})} \cdot \frac{\partial \lambda_i(x_c,\tilde{F})}{\partial x_c} \right)^2
\right) + \mathbb{E}_{I_1,...,I_n} \left( \sum_i\sum_{j \neq i}
\frac{I_i \cdot I_j }{\lambda_i(x_c,\tilde{F}) \cdot \lambda_j(x_c,\tilde{F})} \cdot \frac{\partial
  \lambda_i(x_c,\tilde{F})}{\partial x_c} \cdot \frac{\partial \lambda_j(x_c,\tilde{F})}{\partial x_c}
\right)\nonumber\\
& -  \sum_i\sum_j \frac{\partial \lambda_i(x_c,\tilde{F})}{\partial x_c} \cdot
\frac{\partial \lambda_j(x_c,\tilde{F})}{\partial x_c} \nonumber\\
= & \sum_i \frac{\left( \lambda_i(x_c,\tilde{F})+ \lambda_i(x_c,\tilde{F})^2
  \right)}{\lambda_i(x_c,\tilde{F})^2} \cdot \left( \frac{\partial
  \lambda_i(x_c,\tilde{F})}{\partial x_c} \right)^2 + \sum_i\sum_{j \neq i}
\frac{\partial \lambda_i(x_c,\tilde{F})}{\partial x_c} \cdot \frac{\partial
  \lambda_j(x_c,\tilde{F})}{\partial x_c}
-  \left( \sum_i \frac{\partial \lambda_i(x_c,\tilde{F})}{\partial x_c} \right) ^2 \nonumber\\
= & \sum_i \frac{1}{\lambda_i(x_c,\tilde{F})} \cdot \left( \frac{\partial
  \lambda_i(x_c,\tilde{F})}{\partial x_c} \right)^2
\end{eqnarray}
where we have used, as in Appendix~\ref{proof_pro_photo_1D}, the facts
that $\mathbb{E}(I_i)= \lambda_i$, $\mathbb{E}(I_i^2)= \lambda_i^2 +
\lambda_i$, and $\mathbb{E}(I_i \cdot I_j) = \mathbb{E}(I_i) \cdot
\mathbb{E}(I_j) = \lambda_i \cdot \lambda_j$, this last expression
since the pixel measurements are un-correlated.
In the case of a Gaussian PSF, it is easy to verify using
equations~(\ref{eq_sub_sec_joint_estimation_1b}),
(\ref{eq_sub_sec_joint_estimation_2}),
(\ref{eq_sub_sec_joint_estimation_4}), and~(\ref{ji}) and replacing
them in equation~(\ref{a11}) that:
\begin{equation}\label{a11g}
\mathcal{I}_{1,1}(n) =  \frac{1}{2 \pi \sigma^2} \cdot \frac{G F^2}{B} \cdot
\displaystyle \sum_{i=1}^{n} \frac{\left( e^{-\gamma(x^-_i-x_c)} -
  e^{-\gamma(x^+_i-x_c)} \right) ^2}{\left( 1 + \frac{1}{\sqrt{2 \pi}
    \, \sigma}\cdot \frac{F}{B} \cdot J_i(x_c) \right) }
\end{equation}

For the cross term we have:
\begin{eqnarray}\label{a12}
\mathcal{I}_{1,2}(n) \equiv & \mathbb{E}_{I_1,...,I_n} \left(
\frac{\partial \ln L(\vec{I};x_c,\tilde{F}) }{\partial x_c} \cdot
\frac{\partial \ln L(\vec{I};x_c,\tilde{F}) }{\partial \tilde{F}}
\right) \nonumber\\
= & \mathbb{E}_{I_1,...,I_n} \left( \sum_i \sum_j \frac{I_i \cdot
  I_j}{\lambda_i(x_c,\tilde{F}) \cdot \lambda_j(x_c,\tilde{F})} \cdot
g_j \cdot \frac{\partial \lambda_i(x_c,\tilde{F})}{\partial x_c} -
\sum_i \sum_j \frac{I_i}{\lambda_i(x_c,\tilde{F})} \cdot g_j \cdot \frac{\partial
    \lambda_i(x_c,\tilde{F})}{\partial x_c} \right) \nonumber\\
& - \mathbb{E}_{I_1,...,I_n} \left(\sum_i \sum_j \frac{I_j}{\lambda_j(x_c,\tilde{F})} \cdot g_j \cdot \frac{\partial
    \lambda_i(x_c,\tilde{F})}{\partial x_c} \right) + \sum_i \sum_j g_j \cdot \frac{\partial
    \lambda_i(x_c,\tilde{F})}{\partial x_c} \nonumber\\
= & \mathbb{E}_{I_1,...,I_n} \left( \sum_i
\frac{I_i^2}{\lambda_i(x_c,\tilde{F})^2} \cdot g_i \cdot
\frac{\partial \lambda_i(x_c,\tilde{F})}{\partial x_c} \right) +
\mathbb{E}_{I_1,...,I_n} \left( \sum_i\sum_{j \neq i} \frac{I_i \cdot
  I_j }{\lambda_i(x_c,\tilde{F}) \cdot \lambda_j(x_c,\tilde{F})} \cdot
g_j \cdot \frac{\partial \lambda_i(x_c,\tilde{F})}{\partial x_c} \right) \nonumber\\
& - \sum_i \sum_j g_j \cdot \frac{\partial \lambda_i(x_c,\tilde{F})}{\partial x_c} \nonumber \\
= & \sum_i \frac{\left( \lambda_i(x_c,\tilde{F})+
  \lambda_i(x_c,\tilde{F})^2 \right)}{\lambda_i(x_c,\tilde{F})^2}
\cdot g_i \cdot  \frac{\partial
  \lambda_i(x_c,\tilde{F})}{\partial x_c} + \sum_i\sum_{j \neq
  i} g_j \cdot \frac{\partial \lambda_i(x_c,\tilde{F})}{\partial x_c}
- \sum_i \sum_j g_j \cdot \frac{\partial \lambda_i(x_c,\tilde{F})}{\partial x_c} \nonumber\\
= & \sum_i \frac{g_i}{\lambda_i(x_c,\tilde{F})} \cdot  \frac{\partial
  \lambda_i(x_c,\tilde{F})}{\partial x_c}
\end{eqnarray}
Under a Gaussian PSF, we can also verify using
equations~(\ref{eq_sub_sec_joint_estimation_1b}),
(\ref{eq_sub_sec_joint_estimation_2}),
(\ref{eq_sub_sec_joint_estimation_4}), and~(\ref{ji}), and replacing
them in equation~(\ref{a12}) that:
\begin{equation}\label{a12g}
\mathcal{I}_{1,2}(n) = \frac{1}{2 \pi \sigma^2} \cdot \frac{F}{B} \cdot
\displaystyle \sum_{i=1}^{n} \frac{\left( e^{-\gamma(x^-_i -x_c)} -
  e^{-\gamma(x^+_i-x_c)} \right ) \cdot J_i(x_c)}{\left( 1 +
  \frac{1}{\sqrt{2 \pi} \, \sigma}\cdot \frac{F}{B} \cdot J_i(x_c)
  \right) }.
\end{equation}
Of course, by symmetry, $\mathcal{I}_{1,2} =
\mathcal{I}_{2,1}$.

Finally, for the last matrix element, one has:
\begin{eqnarray}\label{a22}
\mathcal{I}_{2,2}(n) \equiv & \mathbb{E}_{I_1,...,I_n} \left( \left(
\frac{\partial \ln L(\vec{I};x_c,\tilde{F}) }{\partial \tilde{F}}
\right)^2 \right) \nonumber\\
=  & \mathbb{E}_{I_1,...,I_n} \left( \sum_i \sum_j
\left( \frac{I_i \cdot I_j}{\lambda_i(x_c,\tilde{F}) \cdot
  \lambda_j(x_c,\tilde{F})} \cdot g_i \cdot g_j - 2 \cdot
\frac{I_i}{\lambda_i(x_c,\tilde{F})} \cdot g_j \right) \right) + 1 \nonumber\\
= & \mathbb{E}_{I_1,...,I_n} \left( \sum_i \left(\frac{I_i
}{\lambda_i(x_c,\tilde{F})} \cdot g_i \right)^2 \right) +
\mathbb{E}_{I_1,...,I_n} \left( \sum_i\sum_{j \neq i} \frac{I_i \cdot
  I_j }{\lambda_i(x_c,\tilde{F}) \cdot \lambda_j(x_c,\tilde{F})} \cdot
g_i \cdot g_j \right)\nonumber -1 \nonumber\\
= & \sum_i \frac{\left( \lambda_i(x_c,\tilde{F})+
  \lambda_i(x_c,\tilde{F})^2 \right)}{\lambda_i(x_c,\tilde{F})^2}
\cdot g_i^2 + \sum_i\sum_{j \neq i} g_i \cdot g_j - 1 \nonumber\\
= & \sum_i \frac{g_i^2}{\lambda_i(x_c,\tilde{F})},
\end{eqnarray}
where we have used the fact that $\sum_i g_i =1$. In the case of a Gaussian PSF, it is easy to verify using equations~(\ref{eq_sub_sec_joint_estimation_1b}),
(\ref{eq_sub_sec_joint_estimation_2}),
(\ref{eq_sub_sec_joint_estimation_4}), and~(\ref{ji}) and replacing
them in equation~(\ref{a22}) that:
\begin{equation}\label{a22g}
\mathcal{I}_{2,2} = \frac{1}{2 \pi \sigma^2} \cdot \frac{1}{GB}
\cdot \displaystyle \sum_{i=1}^{n} \frac{J_i(x_c)^2}{\left( 1 +
  \frac{1}{\sqrt{2 \pi} \, \sigma}\cdot \frac{F}{B} \cdot J_i(x_c)
  \right)}.
\end{equation}

To conclude, the inverse of the Fisher matrix, which is what we require to
obtain the \crra\ bound, would thus be given by:
\begin{equation}
\mathcal{I}(n)^{-1} = \frac{1}{\Delta} \cdot
\left(
\begin{array}{rr}
\mathcal{I}_{22} & - \mathcal{I}_{12} \\
- \mathcal{I}_{12} &  \mathcal{I}_{11}
\end{array}
\right)
\end{equation}
where $\Delta = \mathcal{I}_{11} \cdot \mathcal{I}_{22} -
\mathcal{I}_{11}^2$ is the determinant of the Fisher matrix.

\newpage

\acknowledgments Rene A. Mendez acknowledges partial support from
project PFB-06 CATA-CONICYT and from project IC120009 "Millennium
Institute of Astrophysics (MAS)" of the Iniciativa Científica Milenio
del Ministerio de Economía, Fomento y Turismo de Chile. Jorge F. Silva
and Rodrigo Lobos acknowledges support from FONDECYT - CONICYT grant
\# 1140840. We would also like to acknowledge several useful comments
from an anonymous referee which lead to a better discussion of
Section~\ref{range} and the introduction of Section~\ref{variable} and
Tables~\ref{tabunder} and~\ref{tabpsfbac}.

\clearpage

\begin{figure}
\centering
\epsscale{.80}
\plotone{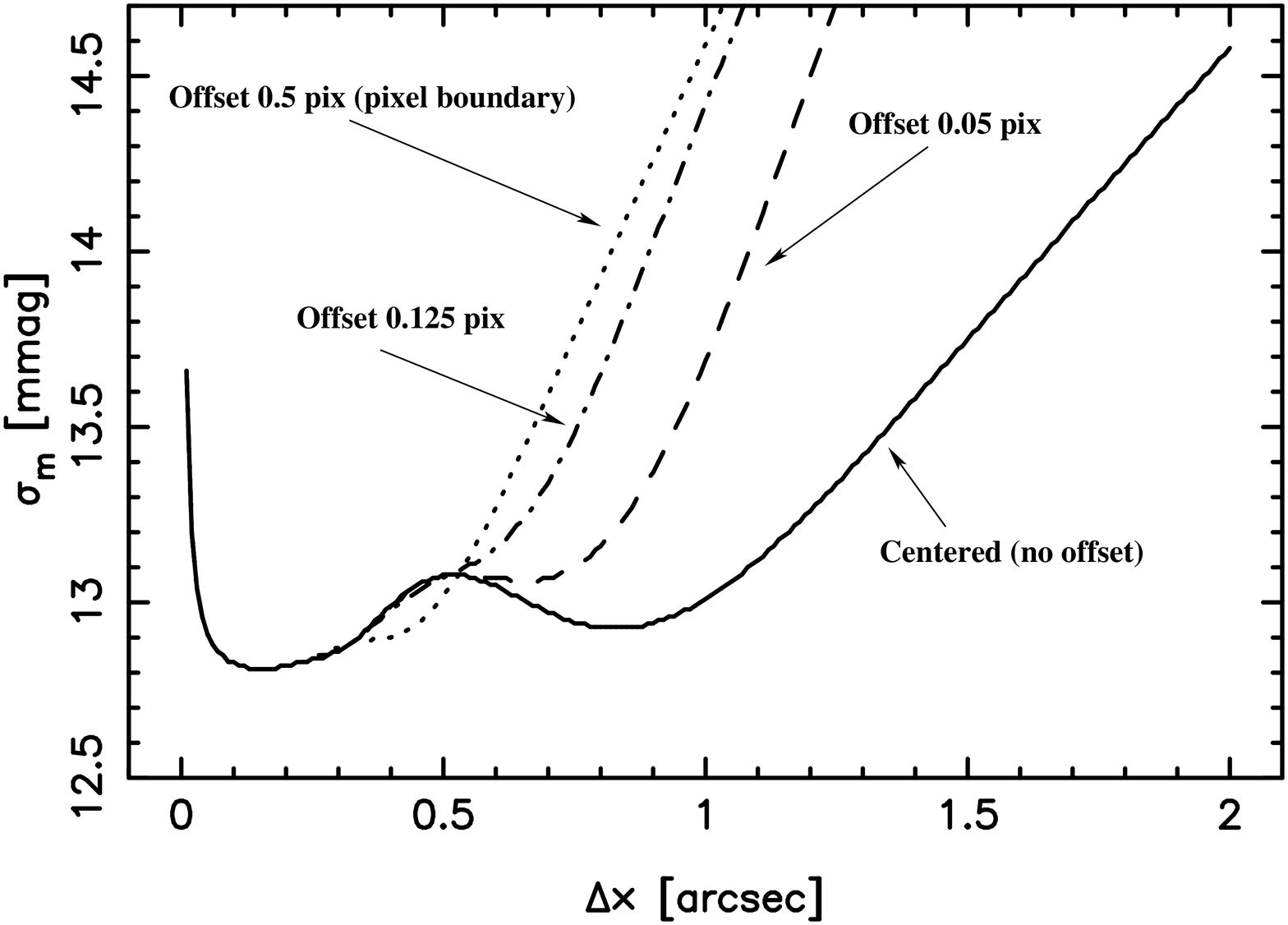}
\caption{Photometric \crra\ bound as given by the Fisher matrix
  coefficients in equation~(\ref{exact}), in milli-magnitudes, as a
  function of detector pixel size $\Delta x$ in arcsec. The curves
  were computed for a detector with $RON=5$~e$^-$, $D=0$~e$^-$,
  $G=2$~e$^-$/ADU, a background of $f_s=2\,000$~ADU/arcsec
  (representative of ground-based observations on moonless nights
  through optical broad-band filters with 600~sec exposure time on a
  good site), and for a Gaussian source with $FWHM=0.5$~arcsec and $F
  = 5\,000$~ADU (corresponding to a $S/N \sim 80$). The solid, dashed,
  dot-dashed, and dotted lines are for sources which are centered,
  off-center by 0.05, 0.125 and 0.5~pix (equal to the pixel boundary)
  respectively. Compare with Figure~1 in \citet{men13}. \label{figpd}}
\end{figure}

\clearpage

\begin{figure}
\centering
\epsscale{.80}
\plotone{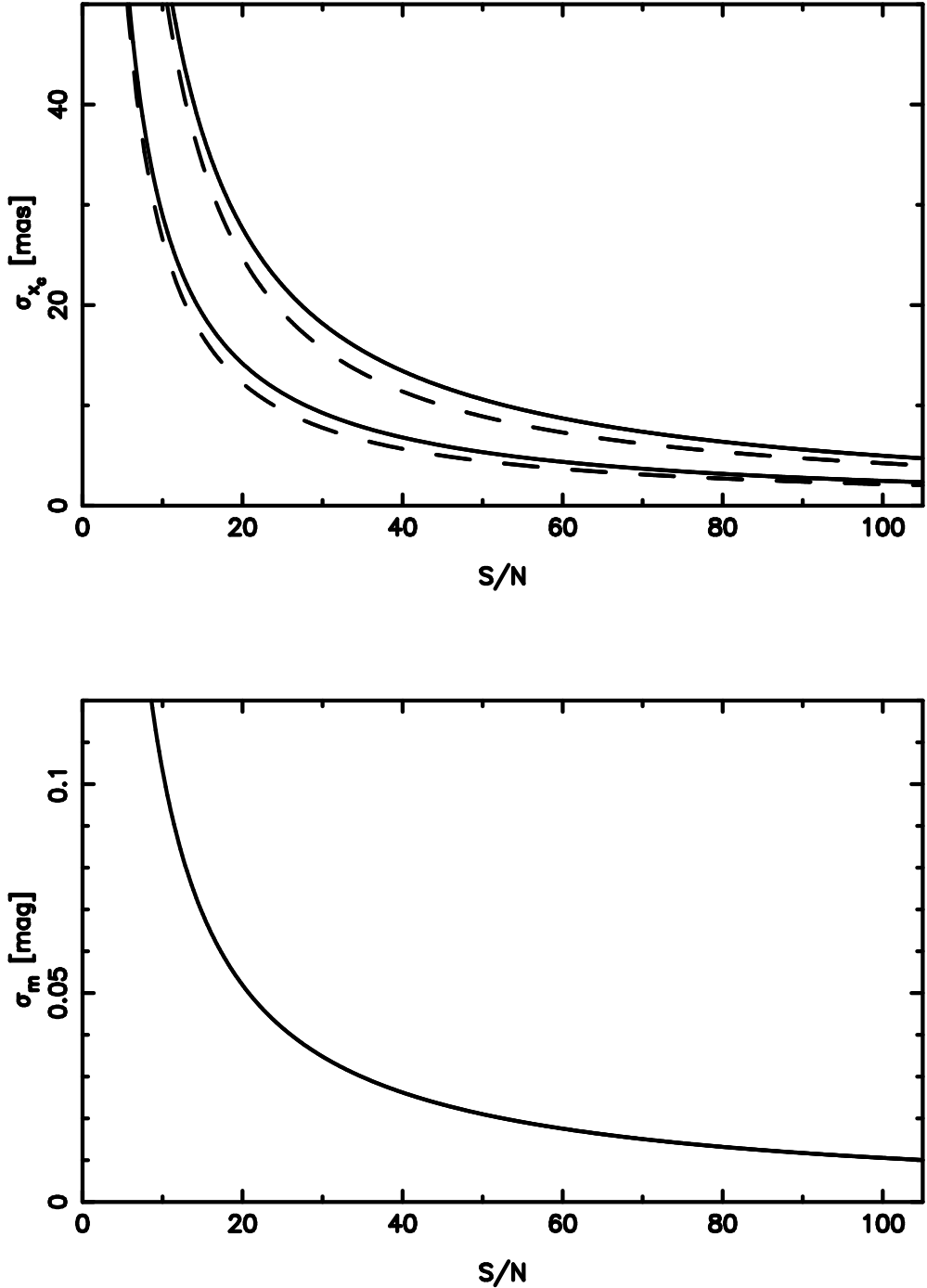}
\caption{Astrometric (top) and photometric (bottom) \crra\ limits as
  given by equation~(\ref{exact}) as a function of $S/N$ for a
  Gaussian source. All the curves were computed for the same detector
  parameters as those of Figures~\ref{figpd}, with $\Delta x =
  0.2$~arcsec and the source centered on a pixel. In the upper figure,
  the solid lines correspond to a $FWHM=1.0$~arcsec (upper line) and
  $FWHM=0.5$~arcsec (lower line), both with $f_s=2\,000$~ADU/arcsec.
  The corresponding dashed lines are the predictions for the same two
  $FWHM$, but with no background ($f_s=0$~ADU/arcsec), the only source
  of noise comes in this case from the readout electronics. In the
  lower panel, the curves for $FWHM=1.0$ and 0.5~arcsec, as well as
  for $f_s=2\,000$ and 0~ADU/arcsec overlap with each other - see text
  for a discussion. \label{figcrap}}
\end{figure}

\clearpage
\begin{figure}
\epsscale{.80}
\centering
\plotone{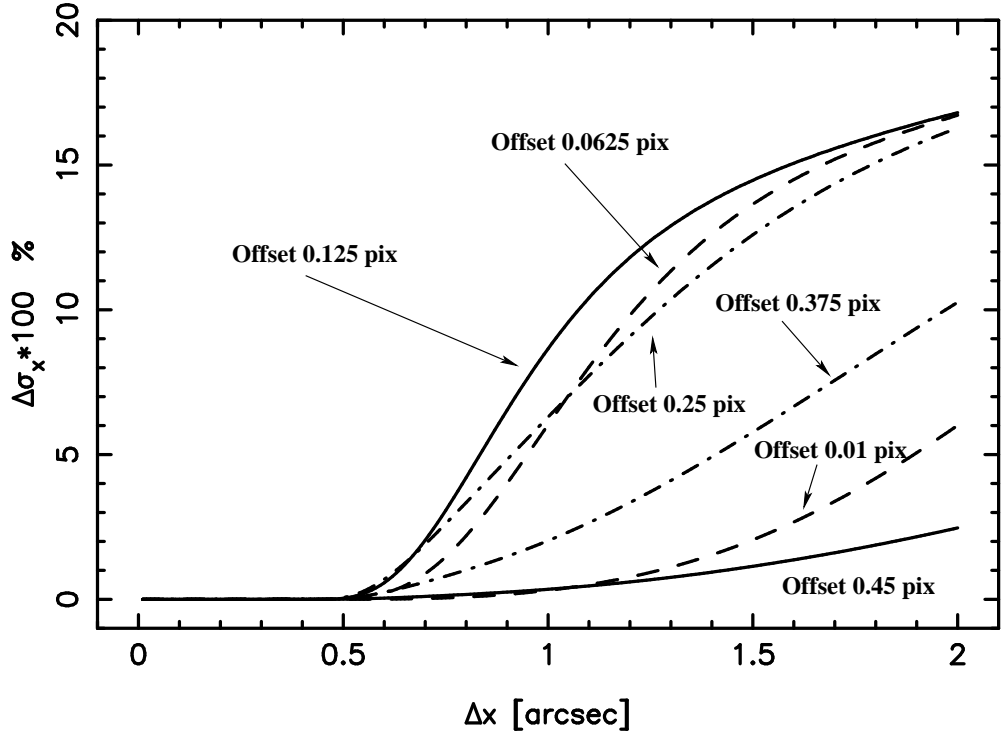}
\caption{Fractional difference in the \crra\ bound computed in 2D and
  1D, as a function of detector pixel size for the same detector
  parameters as those of Figures~\ref{figpd}, and for a source with
  $FWHM=0.5$~arcsec, $F=5\,000$~ADU, and $f_s=2\,000$~ADU/arcsec ($S/N
  \sim 80$). The upper (lower) solid line is for a pixel offset of
  0.125~pix (0.45~pix), the upper (lower) dashed line is for a pixel
  offset of 0.0625~pix (0.01~pix), and the upper (lower) dot-dashed
  lines is for a pixel offset of 0.25~pix (0.375~pix)
  respectively. Note: The pairing of line types is done only to avoid
  crowding of the figure.\label{figdiffa}}
\end{figure}

\clearpage

\begin{deluxetable}{cc|cc|cc|cc}
\tabletypesize{\scriptsize} \rotate \tablecaption{Effect of
  under-sampling, pixel offset, and $S/N$ on the
  \crra\ limit.\label{tabunder}} \tablewidth{0pt} \tablehead{
  \multicolumn{2}{c}{} & \multicolumn{2}{c}{$S/N=5$} &
  \multicolumn{2}{c}{$S/N=30$} & \multicolumn{2}{c}{$S/N=100$}
  \\ $\Delta x$ & Offset & $\sigma_{{x}_c}$, $\sigma_{{x}_{c_{1D}}}$ &
  $\sigma_{{\tilde{F}}}$, $\sigma_{{\tilde{F}}_{1D}}$ &
  $\sigma_{{x}_c}$, $\sigma_{{x}_{c_{1D}}}$ & $\sigma_{{\tilde{F}}}$,
  $\sigma_{{\tilde{F}}_{1D}}$ & $\sigma_{{x}_c}$,
  $\sigma_{{x}_{c_{1D}}}$ & $\sigma_{{\tilde{F}}}$,
  $\sigma_{{\tilde{F}}_{1D}}$ \\ arcsec & pix & mas & \% & mas & \% &
  mas & \% } \startdata 0.25 & 0.125 & 54 , 54 & 18, 18 & 8.7, 8.7 &
3, 3 & 2.5, 2.5 & 1, 1 \\ 0.5 & 0.125 & 87 , 85 & 18, 18 & 13, 13 & 3,
3 & 3.2, 3.2 & 1, 1 \\ 1.0 & 0.125 & 443, 341 & 26, 20 & 60, 50 & 4,
3.5 & 12, 11 & 1, 1 \\ 1.5 & 0.125 & 3930, 2831 & 32, 23 & 529, 408 &
4.6, 3.6 & 102, 89 & 1, 1 \\ \hline 0.25 & 0.25 & 54 , 54 & 18, 18 &
8.7, 8.7 & 3, 3 & 2.5, 2.5 & 1, 1 \\ 0.5 & 0.25 & 69 , 67 & 19, 18 &
11, 11 & 3, 3 & 3.0, 3.0 & 1, 1 \\ 1.0 & 0.25 & 173, 140 & 26, 21 &
24, 21 & 4, 3.5 & 5.3, 5.1 & 1, 1 \\ 1.5 & 0.25 & 541, 403 & 32, 24 &
73, 58 & 4, 3 & 15, 13 & 1, 1 \\ \enddata \tablecomments{All
  \crra\ estimates used a detector with $G=2$~e$^-$/ADU, $RON=5$~e$^-$
  and no dark noise, a background of $f_s=2\,000$~ADU/arcsec, and a
  source with a $FWHM=0.5$~arcsec (same values as for
  Figure~\ref{figdiffa}). The upper part of the table is for a pixel
  offset of 0.125~pix, while the lower part is for a pixel offset of
  0.25~pix, see text for details.}
\end{deluxetable}

\clearpage

\begin{deluxetable}{cc|cc|cc|cc}
\tabletypesize{\scriptsize}
\rotate
\tablecaption{Effect of changes in the $FWHM$ and the background on the
  \crra\ limit of under-sampled images.\label{tabpsfbac}}
\tablewidth{0pt}
\tablehead{
\multicolumn{2}{c}{} & \multicolumn{2}{c}{$S/N=5$} & \multicolumn{2}{c}{$S/N=30$} &
\multicolumn{2}{c}{$S/N=100$} \\
$FWHM$ & $f_s$ & $\sigma_{{x}_c}$ & $\sigma_{{\tilde{F}}}$ &
$\sigma_{{x}_c}$ & $\sigma_{{\tilde{F}}}$ & $\sigma_{{x}_c}$ &
$\sigma_{{\tilde{F}}}$ \\
arcsec & ADU arcsec$^{-1}$ & mas & \% & mas & \% & mas & \%
}
\startdata
0.5 & 2000 & 1092 & 19 & 147 & 3 & 29 & 1 \\
0.6 & 2000 &  562 & 20 &  76 & 3 & 15 & 1 \\
\hline
0.5 & 2200 & 1145 & 20 & 155 & 3 & 30 & 1 \\
0.6 & 2200 &  589 & 21 &  80 & 3 & 16 & 1 \\
\hline
\hline
0.5 & 2000 &  70 & 26 & 11 & 4 & 3 & 1 \\
0.6 & 2000 &  84 & 26 & 13 & 4 & 4 & 1 \\
\hline
0.5 & 2200 & 73 & 27 & 11 & 4 & 3 & 1 \\
0.6 & 2200 & 88 & 27 & 13 & 4 & 4 & 1 \\

\enddata \tablecomments{All \crra\ estimates used a detector with
  $G=2$~e$^-$/ADU, $RON=5$~e$^-$ no dark noise, and $\Delta x =
  1.0$~arcsec, a background of $f_s=2\,000$~ADU/arcsec, and a source
  with a $FWHM=0.5$~arcsec (same values as for
  Figure~\ref{figdiffa}). The upper part of the table is for a source
  with no offset (i.e., centered on a pixel, worst case for
  astrometry, best case for photometry), whereas the lower part is for
  a source at a pixel boundary (best case for astrometry, worst case
  for photometry).}
\end{deluxetable}

\end{document}